\documentclass[conference]{IEEEtran}
\usepackage{graphicx}
\usepackage{amsmath}
\usepackage{listings}
\usepackage{url}
\usepackage[rightcaption]{sidecap}
\usepackage{epstopdf}
\usepackage[caption=false]{subfig}
\usepackage{multirow}
\usepackage{wrapfig}
\usepackage{array}
\usepackage[ruled,vlined,linesnumbered]{algorithm2e}
\usepackage{tikz}
\usepackage[shortlabels]{enumitem}
\usepackage{amsthm}
\theoremstyle{definition}
\newtheorem{definition}{Definition}[]

\ifCLASSINFOpdf
\else
\fi

\begin{document}
%


\title{Exploiting RapidWright in the Automatic Generation of Application-Specific FPGA Overlays}


%




\author{
\IEEEauthorblockN{
Joel Mandebi Mbongue\IEEEauthorrefmark{1},
Danielle Tchuinkou Kwadjo\IEEEauthorrefmark{1}, 
Christophe Bobda\IEEEauthorrefmark{1}}

\IEEEauthorblockA{\IEEEauthorrefmark{1}ECE Department, University of Florida, Gainesville FL, USA\\
Email: jmandebimbongue@ufl.edu, dtchuinkoukwadjo@ufl.edu, cbobda@ece.ufl.edu}
}


\maketitle

\begin{abstract}

 Overlay architectures implemented on FPGA devices have been proposed as a means to increase FPGA adoption in general-purpose computing. They provide the benefits of software such as flexibility and programmability, thus making it easier to build dedicated compilers. However, existing overlays are generic, resource and power hungry with performance usually an order of magnitude lower than bare metal implementations. As a result, FPGA overlays have been confined to research and some niche applications. In this paper, we introduce \textit{Application-Specific FPGA Overlays} (AS-Overlays), which can provide bare-metal performance to FPGA overlays, thus opening doors for broader adoption. Our approach is based on the automatic extraction of hardware kernels from data flow applications. Extracted kernels are then leveraged for application-specific generation of hardware accelerators. Reconfiguration of the overlay is done with RapidWright which allows to bypass the HDL design flow. Through prototyping, we demonstrated the viability and relevance of our approach. Experiments show a productivity improvement up to 20$\times$ compared to the state of the art FPGA overlays, while achieving  over 1.33$\times$ higher Fmax than direct FPGA implementation and the possibility of lower resource and power consumption compared to bare metal.
\end{abstract}


%
\IEEEpeerreviewmaketitle

\section{Introduction}

Over the past decades, FPGAs have continuously matured  and now contain millions of logic gates, thousands of DSP blocks, megabytes of BRAMs, and other types of resources. This development opens doors to unprecedented hardware acceleration in several computing domains such as deep learning, image and scientific processing, and cloud computing. For instance, Xilinx recently released the U$250$ Alveo card powered by UltraScale+ FPGAs for data center and artificial intelligence acceleration. The U$250$ gathers four super logic regions each containing approximately $340000$ logic elements, $20$MB of BRAM, $90$MB of UltraRAM, and $3000$ DSP slices \cite{alveo}\cite{u250}.
The Intel Arria 10 in Microsoft Cloud delivers about $1.1$ million logic elements, $3036$ DSP logics, and $67$MB of BRAM \cite{microsoftCloud}\cite{arria10}. Nevertheless, these feature improvements have not translated into widespread use of FPGAs. One reason is that designing for FPGAs remains a challenging endeavour including required hardware expertise and long compilation time, which limits the efficient use of FPGA accelerators to niche disciplines involving highly skilled hardware engineers.

To help addressing that limitation, High Level Synthesis (HLS) have been proposed \cite{winterstein2013high}\cite{czajkowski2012opencl}. It focuses on high-level functionality rather than low level implementation. However, the hardware expertise and the prohibitive compilation times (especially placement and routing) still limit productivity and mainstream adoption.  The need to make FPGAs more accessible to application developers who are accustomed to software API abstractions
and fast development cycles therefore remains. 

FPGA overlays have been developed to promote FPGAs to a wider user community and for increased design productivity. In general, overlays use coarse-grained processors, which can be programmed from a function call, in a $2$D intercommunication infrastructure that allows parallel processing and data exchange among the processors \cite{mandebi2018flexitask}\cite{brant2012zuma}\cite{hartenstein2001coarse}\cite{li2016area}\cite{Metzner2015}. The software nature of the coarse-grained processors makes it possible to develop efficient compilers for automatic mapping of sequential applications, thus increasing their acceptance in the software community. 

Unfortunately, these advantages come at the cost of area and performance, limiting overlays to relatively small to moderate applications. Indeed, FPGA overlays are usually an order of magnitude slower than bare metal implementations, and consume way more resource and power. Because the main purpose of FPGAs is hardware acceleration, overlays have therefore not been able to breakthrough.  

In this work, we introduce \textit{Application-Specific FPGA Overlays} (AS-Overlays), a novel form of FPGA overlays designed for Data Flow Applications. AS-Overlays provide the flexibility of state-of-the-art overlays on one hand and bare metal performance on the other hand. It leverage application specific architectural components for efficient bare metal implementation of functions needed by run-time applications, effectively eliminating the intermediate layers of conventional overlays. We propose an approach for automatic generation of overlay kernels from applications. The proposed approach differs from traditional HLS in that kernels are identified from a set of high-level programming language (HLPL) applications, with no hardware description language (HDL) generation and no usage of domain-specific language (DSL). Specifically, our contribution includes :
\begin{enumerate}[(1)]

\item An application-specific FPGA overlay generation flow for productivity, performance, and power consumption improvement. 


\item An automatic identification of application kernels through intermediate representation inspection using the Low Level Virtual Machine (LLVM) \cite{lattner2004llvm}, a compilation and code instrumentation framework.

\item Systematic hardware generation from identified kernels using RapidWright to shorten design cycles and generate tailored netlists \cite{lavin2018rapidwright}.


\end{enumerate}
In the rest of the paper, section \ref{lab:relatedWork} revisits recent research, section \ref{lab:designFlow} describes our proposed AS-Overlays design flow, section \ref{sec:dfg_app} discusses RapidWright features and defines data flow application in the context of this work, section \ref{sec:kernelMining} details kernels mining within applications, section \ref{sec:hardwareGen} discusses systematic hardware generation, section \ref{lab:experimentalObservation} presents experimental observations, and section \ref{lab:conclusion} concludes the paper.

\section{Related Work}\label{relwork}
\label{lab:relatedWork}
Published work in coarse-grained reconfigurable architectures and FPGA overlays such as 
\cite{mirsky1996matrix, Kapre2015, brant2012zuma} are essentially \textit{dataflow machines}, usually consisting of small arithmetic and logic units, registers, all of which are immersed in an switch-based interconnect structure. The processors are homogeneous and programmable, and not tailored for specific applications. Overlays such as Hoplite \cite{Kapre2015}, FLexiTASK \cite{mandebi2018flexitask},  and Quattor \cite{Metzner2015} have dedicated optimization, mostly focusing on the interconnect and communication infrastructure. The Hoplite-DSP \cite{Chethan2016} is the closest to the approach proposed in this work in that it leverages DSP blocks on the FPGA for dedicated implementation. However, Hoplite-DSP is still a generic architecture with homogeneous processing units. 

Several research in the literature have discussed solutions for automating the generation of hardware accelerators on FPGAs. Ma et al. \cite{ma2015run} proposed a flow relying on pre-synthesized functions for runtime generation of FPGA accelerators. It nevertheless requires mastering a specific DSL and is not optimized for performance. Ishebabi et al. \cite{ISHEBABI200963} presented methods for automatic synthesis targeting arrays of multiprocessors on chip using exact formulations such as integer linear programming of answer set programming. However, the search for kernel is done using profiling. In the same line of idea, Koeplinger et al. \cite{koeplinger2016automatic} present a framework for automatic generation of efficient application specific FPGA accelerators. Parallel pattern inputs aim to raise the level of abstraction of programmers in addition to providing purposeful information to the compiler. Their approach nevertheless relies on parallel inputs which do not necessarily reflect how developers would typically implement applications. 
Other tools such as LegUp \cite{canis2013legup} and Vivado HLS \cite{winterstein2013high} allow designers to write code in HLPL and then compile to a register transfer level (RTL) design specification. Though Vivado HLS can deliver competitive quality of results (QoR) compared to manual RTL \cite{cong2011high}, it still incurs design efforts and long compilation time. LegUp provides a built-in profiler to identify computation intensive code regions for acceleration. Applications are then modified to run partially on MIPS CPU and hardware accelerator on FPGA. Runtime communication between CPU and accelerator coupled with potential cache coherency issues might limit performances achievable by the platform.
The work of Cong et al. \cite{cong2008pattern} is similar to ours in that they applied graph-based techniques to identify frequent patterns by analyzing graph edit distances. That work nevertheless differs from ours as patterns are detected for optimized FPGA resource sharing during the binding in behavioral synthesis. 


In contrast to the aforementioned research, our AS-Overlays generation flow leverages advanced graph mining techniques to find kernels in applications and builds a library of accelerators that can be combined into an FPGA overlay to improve performance, productivity, and power consumption.

%

\section{Design Flow}
\label{lab:designFlow}
\begin{figure}[]
\begin{center}
   \includegraphics[width=0.45\textwidth]{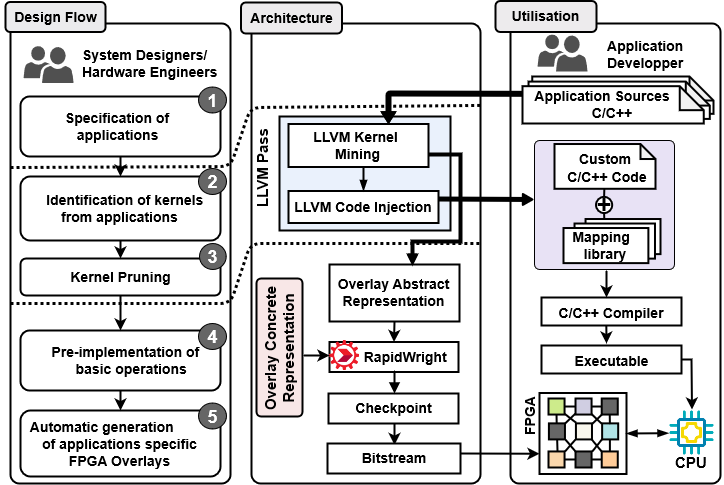}
\end{center}
   \caption{Design Flow, Utilization Flow and Architecture}
\label{fig:architecture}
\end{figure}

The major limitation of FPGA overlays resides in that they most often feature more resources than what is actually needed, resulting in increased power consumption and performance lost. They are regularly made of several processing elements (PEs) and interconnect. PEs generally contain some registers and a functional unit capable of executing a set of functions, resulting in architectures not optimized for specific tasks. In Figure \ref{fig:architecture}, we propose a \textit{Design Flow} to build FPGA overlays that can compete with bare metal implementations. Few steps are necessary to produce an AS-Overlay: 
\begin{enumerate}
    \item Specify the application with a HLPL.
    \item Inspect the bytecode or intermediate representation (IR) of the application at compile time to extract compute-intensive code sections that we identify as kernels.
    \item Optimize kernels to remove unneeded instructions.
    \item Manually pre-synthesize basic operations from the IR instruction set using vendor tools: this step is done exactly once, and the synthesized netlists can be reuse in several other applications.
    \item Combine the pre-synthesized basic operations according to kernel descriptions to generate hardware circuits.
\end{enumerate}
Figure \ref{fig:architecture} also illustrates the overall \textit{Architecture} supporting the proposed design flow. After the identification of kernels, the LLVM Pass generates a new source code equivalent to the input program, in which kernels' instructions are replaced by hardware calls.

We use LLVM to search for kernels as it allows transparent optimization on applications written in arbitrary HLPL. Each application is parsed with an LLVM Frontend to output an IR. The produced LLVM IR is then converted into data flow graphs ("Overlay Abstract Representation" in Figure \ref{fig:architecture}) for analysis, and kernels are identified. RapidWright is further leveraged to automatically generate kernel netlists by assembling as in a puzzle, a set of pre-synthesized LLVM IR operations. We use RapidWright because it is designed to quickly stitch together pre-implemented modules with minimal QoR loss. Finally, hardware kernels are embedded within PEs of an arbitrary overlay architecture ("Overlay Concrete Representation" in Figure \ref{fig:architecture}), and Vivado is used to place and route the AS-Overlay. In the \textit{Utilization} flow, a new optimized C/C++ code alongside the mapping library can now be compiled and run on a SoC. The mapping library is made of a set of functions handling data copy to/from the FPGA, removing the need for hardware expertise. In the rest of the paper, the focus is mainly set on kernel mining and hardware generation.

 
 \section{Preliminary}
\begin{figure}[h]
\centering
\vspace{-10pt}
\includegraphics[width=0.32\textwidth]{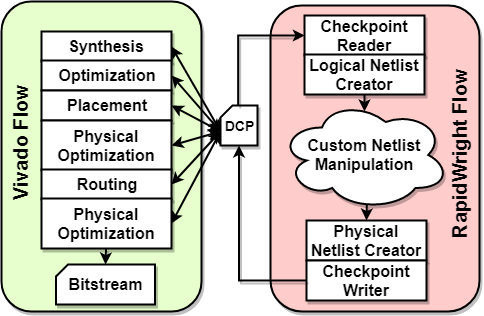}
\caption{\label{fig:rapidwrightFlow} RapidWright Flow vs Vivado Flow.}
\vspace{-8pt}
\end{figure}

 \textbf{RapidWright} \cite{lavin2018rapidwright} is an open source Java framework from Xilinx Research Labs that provides a bridge to Vivado back-end at different compilation stages through design checkpoint (DCP) files. By making available logical/physical netlist data structures and functions, it enables custom netlist manipulation and direct access to logic and routing resources such as look-up tables (LUT), flip-flops (FF) and programmable interconnect points from a Java API (see Figure \ref{fig:rapidwrightFlow}). 
As opposed to vendor tools that are closed source, we believe the full access to RapidWright internal features and design resources makes it suitable for design flow exploration and the implementation of targeted FPGA solutions.
 
\textbf{Data Flow Applications:}
\label{sec:dfg_app}
the data flow concept refers to a way of looking at the execution flow of instructions in an application. It gives a perspective on operations and their interactions. For modeling purposes, we use the data flow graph (DFG) representation in which nodes represent machine operations, internal edges, data flowing between pairs of operations, and external edges, connections with inputs/outputs.
We study these applications because they represent the base of calculation in several computing domains such as image and video processing, or deep learning. In the subsequent sections, we will study how kernels are identified and corresponding hardware is generated. In the rest of the paper, we will refer to "data flow applications" as "applications".
 
\section{Kernel Mining}
\label{sec:kernelMining}
This section discusses how compute-intensive code portions are extracted from the IR of applications. We begin with background definitions necessary to understand terminologies, then we detail data structures and the kernel mining algorithm.
\subsection{Background Definitions}
\begin{definition}
A \textit{Control Data Flow Graph} (CDFG) is a directed graph $G = (V, E, L, l)$, where $V$ represents the set of vertices, $E \subseteq V \times V$ the set of edges, and $L$ the set of labels, with $l :V \cup E \to L $ being the labeling of vertices and edges. In the context of this work, a \textit{graph} is a CDFG defined at a basic block (BB) level. 
\end{definition}
\begin{definition}
An \textit{isomorphism} between a graph G and a graph H is a bijective function:
$f:V(G) \to V(H) 
l_G (u) = l_G (f(u)),\; \; for \; u \to V(G) 
(f(u), f(v)) \in E(H), \; \;
l_G (u, v) = l_G (f(u), f(v)), \; \; for \; u, v \to E(G) 
$. It measures the similarity between G and H, and therefore prevents recording multiple instances of the same graph when a function \textit{f} can be found.
A \textit{subgraph isomorphism} from G to H is an isomorphism from G to subgraph H.
\end{definition}
\begin{definition} Let $GS$ be a set of CDFGs. The \textit{support} is the minimum frequency of appearance of a subgraph $g$ in $GS$.
\end{definition}
\begin{definition} Given a set of CDFGs $GS=\{g_i | i=1..n\}$ and a threshold $\alpha$ (in this case the minimum support value), the \textit{kernel mining} consists in finding graphs $g$ in $GS$, such that $support(g) >= \alpha$. \textit{Kernels} will then represent the set of graphs $g$.
\end{definition}

\begin{definition}
A Depth First Search (DFS) traversal of a graph defines the order in which its edges are visited: that sequence of edges represent the\textit{ DFS code} of the graph.
\end{definition}

\subsection{Kernel Mining}
Prior to kernel mining, we build a DFG with properties that best capture the input program. Each vertex is labeled with the operation of the instruction it represents, while edges display the order of precedence between operations. The mining algorithm is summarized in Algorithm \ref{algo:kernelDetectionAlgo}.
\begin{algorithm}
\scriptsize
\SetAlgoLined
\SetKwInOut{Input}{Input}\SetKwInOut{Output}{Output}
\Input{LLVM IR, minSup}
\Output {C++ source file}
\BlankLine
\SetKwFunction{FkerneMining}{kernelMining}
\SetKwProg{Pn}{Function}{:}{\KwRet}
  \Pn{\FkerneMining(GS, Fsubgraphs, minSup)}{
  	sort labels of the vertices and edges in GS by frequency (using DFS code)\;
  	remove infrequent vertices and edges\;
  	relabel the remaining vertices and edges (descending)\;
  	S1 := all frequent 1-edge graphs\;
  	sort S1 in DFS lexicographic order\;
  	Fsubgraphs := S1\;
   \ForEach{edge e in S1}{
  	init g with e\;
  	set g.DS=\{$ h | h \in $ GS, e $\in$ E(h)\} \;
  	subgraphMining(GS,Fsubgraphs,g)\;
  	GS := GS - e\;
  	\If{( $ \mid GS \mid < $ minSup)} {
  		break\;
  	}
  	}
 }
 
 \SetKwFunction{FcodeInjection}{codeInjection}
\SetKwProg{Pn}{Function}{:}{\KwRet}
  \Pn{\FcodeInjection(Fsubgraphs)}{
   setLines lines\;
   setVariables var\;
   set files\;
   \ForEach{Graph graph in Fsubgraphs}{
  	\ForEach{Instruction inst in graph}{
  	lines := getInstructionLine(inst)\;
  	var := getVariables(inst)\;
    files := getfileName(inst)\;
  	}
  	injectFunctions()\;
  	}
 }
 
\BlankLine
 \tcc{Main function defined as a ModulePass}
\SetKwFunction{FgenerateDFG}{generateDFG}
\SetKwProg{Pn}{Function}{:}{\KwRet}
  \Pn{\FgenerateDFG(Module M)}{
	setGraphs GS\;
	setGraphs Fsubgraphs\;
  \ForEach{Function FF in  M}{
  	\ForEach{BasicBlock BB in FF}{
  		Graph BBgraph\; 	
  	 \ForEach{Instruction II in bb}{
  	 	BBgraph := II; 
  	 	
  	 }
  		GS := BBgraph;
  }
    \FkerneMining(GS, Fsubgraphs, minSup)\;
    \FcodeInjection(Fsubgraphs)\;
  }
  }
 
\caption{LLVM Pass for Kernel Mining}
\label{algo:kernelDetectionAlgo}
\end{algorithm}

In the control-flow kernel mining, a CDFG normally consists of several DFGs. Nevertheless, the labeling is done such that all the DFGs belonging to a BB remain in the same hierarchy (line 24-26), even if no common edge exists between them. 
Overall, the kernel mining follows several steps among which: 
\begin{enumerate}
    \item \textit{Generate candidates} using a DFS-based approach (line 2),
    \item \textit{Prune the candidates} to remove infrequent vertices and edges (line 3 to 7),
    \item \textit{Evaluate the support value} to decide whether a candidate is a kernel or not (line 8 to 16).
\end{enumerate}

However, the isomorphism search during candidate pruning is known to be NP-complete \cite{cordella2004sub}, and several subgraphs isomorphism techniques as the ones described in \cite{ketkar2005subdue} and \cite{zaki2005efficiently} lead to high computation overhead. One way to mitigate that high overhead consists in computing the canonical form of graphs \cite{miyazaki1997complexity}: if the canonical form of two graphs is identical, the graphs are considered isomorphic. We therefore construct canonical form of DFS codes as in \cite{yan2002gspan}, which the minimum code that can be derived from a graph $g$. Specifically, the strategy consists in:
\begin{enumerate}[(1)]
\item Building frequent subgraphs bottom-up, using DFS code as regularized representation. 
\item Eliminating redundancies via minimal canonical DFS code based on lexicographic ordering. 
\end{enumerate}

 Given that there is a considerable amount of DFS codes, we build a DFS code tree using a lexicographic ordering \cite{yan2002gspan} between DFS codes as follows: A DFS code $a = (a_0, a_1, ..., a_m)$ is parent of DFS code $ b = (a_0, a_1, ..., a_m, b)$ and $b$ is child of $a$  if: (1) each node represents DFS code. (2) The relationships between parents and children conform to the lexicographic ordering. (3) The siblings are consistent with DFS lexicographic order.

The DFS code tree structure is particularly useful in the kernel mining as it allows to make the following  two assumptions:
(1) If a DFS code $\gamma$ is frequent, then every ancestor of $\gamma$ is frequent. (2) If $\gamma$ is not frequent then every descendant of $\gamma$ is not frequent (line 11).

Finally, the custom C/C++ code is generated by replacing kernels' instructions by high-level  functions for hardware acceleration. Original names of the variables and their location are retrieved by inspecting the debug metadata (line 23 - 27) attached to each instruction in the IR. LLVM uses the DWARF \cite{dwarf_home} standardized debugging data format like several other compilers and debuggers to support source layer debugging. The metadata provides the relationships between the generated code and the source code of the original program. 

\begin{wrapfigure}[9]{r}{0.14\textwidth}
\centering
\vspace{-10pt}
\includegraphics[width=0.1\textwidth]{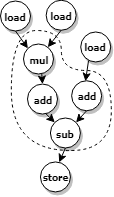}
\vspace{-5pt}
\caption{\label{fig:pruning} Graph Pruning}
\end{wrapfigure}

Once kernels are actually extracted from an application, there is still a need to undergo a final graph pruning. It consists in selecting operations that can actually be mapped on FPGAs. This pruning follows two main stages: \textbf{(1) Removing Load/Store:} codes initially being written for Von Neumann architectures, LLVM IR introduces a set of \textit{load} and \textit{store} instructions that are not needed on FPGA. We therefore only consider operations different from such instructions as illustrated in Figure \ref{fig:pruning}.
 
\textbf{(2) Avoiding Conversions}: LLVM often inserts  casting operations like \textit{\textbf{zext}} that are not qualified for FPGA acceleration.

\begin{figure}[h]
\centering
\vspace{-10pt}
\begin{minipage}[t]{4cm}
  \centering
  \includegraphics[scale=0.21]{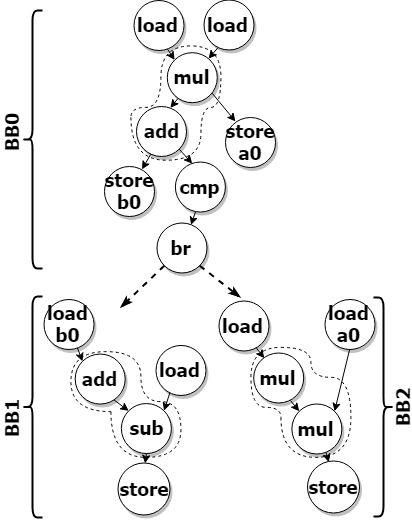}
  \caption{Kernels with Dependencies}
  \label{fig:inter-kernel}
\end{minipage}
\hspace{0.2cm}
\begin{minipage}[t]{4cm}
  \centering
  \includegraphics[scale=0.39]{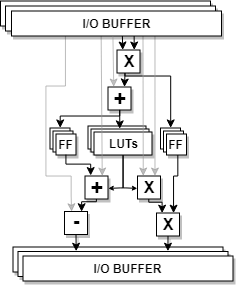}
  \caption{Merged Kernel}
  \label{fig:inter-kernel-circuit}
\end{minipage}
\vspace{-5pt}
\end{figure}

We further study dependencies between basic blocks, searching for additional optimization possibilities. We mainly seek to merge kernels displaying dependencies to save resources and reduce the global latency. As example, Figure \ref{fig:inter-kernel} pictures three kernels spread across basic blocks BB0, BB1, and BB2 (the kernels are encircled with dotted lines).  
Because of the data dependencies between BB0-BB1 (content of the register \textit{b0}) and BB0-BB2 (content of register \textit{a0}), we generate the more complex kernel illustrated in Figure \ref{fig:inter-kernel-circuit}. We insert demuxes for conditional branches and FFs for temporary storage. Instead of having the kernels deployed over three PEs, we can then use a single processing core.

The following section describes how a placed and routed AS-Overlay is obtained from LLVM kernels.
\section{Hardware Generation of Kernels}
\label{sec:hardwareGen}
Initially, the function implemented by a PE in the overlay layout is defined as a 
black-box. We leverage the pre-implemented design flow of RapidWright \cite{lavin2018rapidwright} to produce netlists from kernels previously identified with LLVM. The first step consists in synthesizing basic operations from LLVM IR out-of-context (OOC) with Vivado to create a \textit{library of Modules}. Modules are built OOC to ensure that I/O buffers and global clocks are not inserted into kernel netlists \cite{ooc}. This stage implies a manual implementation (through HDL or HLS) of operations to combine into kernels by an engineer, with the advantage that this step is done once. In the following stage, an application built on the RapidWright API stitches pre-implemented modules following the LLVM kernel DFG descriptions. The hardware kernels thus generated are returned as design checkpoint files and define the functions to be executed in PEs. Finally, the RapidWright application opens the netlist of the overlay (EDIF or DCP files), browses through the design cells,  and reads-in kernel DCPs into PE black-boxes.
\begin{figure}[h]
\centering
\vspace{-10pt}
\includegraphics[width=0.3\textwidth]{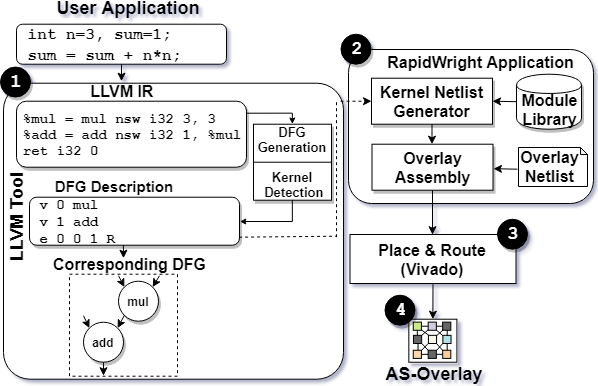}
\caption{\label{fig:applicationOverlay} Example of Overlay Generation.}
\vspace{-10pt}
\end{figure}
Figure \ref{fig:applicationOverlay} illustrates AS-Overlays generation steps. From an application, the LLVM tool identifies kernels and generates corresponding DFGs. Each DFG is dumped into a list of vertices and edges. Vertices start with the character \textit{v}, and are characterized by an identifier and a label denoting the operation. Edges are introduced by a letter \textit{e}, and are defined with an identifier, the two vertices it connects, and a letter (\textit{L} for left, and \textit{R} for right) specifying if the edge gets into the sink vertex through the left or right input. In the next step, a DCP/EDIF containing the layout of the overlay (with the PE functions still being black-boxes) is opened within the RapidWright application, and the generated hardware kernels are successively read-in in PEs, and a new DCP is created for the overlay, this time with each PE implementing a specific function derived from the LLVM kernel mining. Finally, placement and routing are run with Vivado, and a bitstream of the overlay is produced for FPGA deployment. As opposed to the traditional RapidWright pre-implemented flow, which implies synthesizing, placing, and routing modules OOC \cite{lavin2018rapidwright}, basic operations from LLVM IR are only synthesized. Undeniably, for accurate Vivado post-routing timing analysis, partition pin constraints must be defined on input ports of OOC modules, with the consequence of attaching pre-implemented modules to specific FPGA regions \cite{ooc}. Since kernel netlists are automatically generated with RapidWright from DFGs, it is not possible to know in advance what FPGA resources will be used and how they will actually be assembled into hardware kernels. We therefore limit the pre-implementation of LLVM IR operations to the synthesis stage.
\subsection{Datapath Regularization}
\label{sub:datapath_regularization}
\begin{figure}[]
	\centering
    \subfloat[ Non-Regularized Datapath]{
		\label{fig:no-regularization}
		\includegraphics[width=0.2\textwidth]{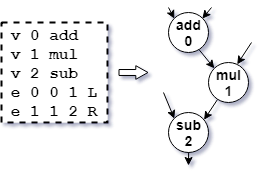}
	}		
	\hspace{-0.2cm}
	\subfloat[Regularized Datapath]{
		\label{fig:regularized}
		\includegraphics[width=0.16\textwidth]{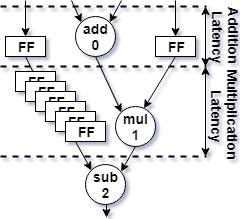}	}
 	\caption{Datapath Regularization}
 	\label{fig:regularization} 
\end{figure}

To reduce overall latency and data management overhead, datapaths must be regularized. Each operation within a kernel  comes with its own latency in number of clock cycles. We must therefore ensure that operands arrive at the boundary of each module at the same time to expect correct results. Figure \ref{fig:no-regularization} shows an example of graph needing regularization. If we assume that addition and multiplication respectively require 1 and 6 clock cycles, the right operand of vertex "\textit{mul 1}" and the left operand of vertex "\textit{sub 2}" must be delayed by 1 and 7 clock cycles. This task is done by the RapidWright application which inserts FFs on the path as illustrated in Figure \ref{fig:regularized}. Inserting FFs do not increase overall latency as the number of FFs is the cumulative latency of operations on the datapath.
\subsection{Processing Elements}
We do not discuss interconnection topology between PEs as the focus is not on obtaining improved/flexible communication; rather, we emphasize architectural features supporting the automatic generation of hardware kernels. 
\begin{wrapfigure}[13]{  r}{0.2\textwidth}
\centering
\vspace{-10pt}
\hspace{-13pt}
\includegraphics[width=0.18\textwidth]{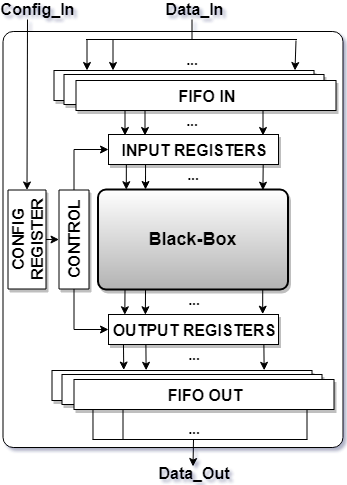}
\caption{\label{fig:pe} PE Architecture}
\end{wrapfigure}
In addition, the proposed AS-Overlay generation flow is designed and well suited for any interconnect topology (mesh, torus, mixed topology, etc \cite{bjerregaard2006survey}) as only the PE processing core will be changed. We therefore look at the minimum architecture set-up that should be embedded in each PE. Figure \ref{fig:pe} illustrates the architecture of PEs. To handle kernels of multiple inputs/outputs as shown in Table \ref{tab:table1}, the I/O buses have parameterizable sizes and are split into 32-bits channels. Inputs and outputs are temporarily stored in I/O queues to avoid data lost in case of multiple clock domains crossing. The \textit{Control} module is configured with the latency of the kernel programmed in the PE, allowing to orchestrate when fetching data from input queues, and when writing results into output queues. The \textit{Black-box} is the core of the PE as it implements one or multiple kernels derived from LLVM code inspection.

\begin{table*}[h]
\scriptsize
\caption {Execution Time Comparison on 3$\times$3 PEs in $\mu$ s } \label{tab:table2}
\resizebox{\textwidth}{!}{%
\begin{tabular}{c|c|c|c|c|c|c|c|c|c|c|c|c|c|c|c}
\hline \hline
\multicolumn{4}{c||}{\textbf{Matrix Mult}} & \multicolumn{4}{c||}{\textbf{Outer Product}} & \multicolumn{4}{c||}{\textbf{Robert Cross}} & \multicolumn{4}{c}{\textbf{Smoothing}} \\ \hline \hline
\multicolumn{1}{c|}{\textbf{Size}}& \multicolumn{1}{c|}{\textbf{\begin{tabular}[c]{@{}c@{}}Bare \\ Metal\end{tabular}}} & \multicolumn{1}{c|}{\textbf{\begin{tabular}[c]{@{}c@{}}Regular \\ Overlay\end{tabular}}} & \multicolumn{1}{c||}{\textbf{\begin{tabular}[c]{@{}c@{}}AS \\ Overlay\end{tabular}}} & \multicolumn{1}{c|}{\textbf{Size}} & \multicolumn{1}{c|}{\textbf{\begin{tabular}[c]{@{}c@{}}Bare \\ Metal\end{tabular}}}& \multicolumn{1}{c|}{\textbf{\begin{tabular}[c]{@{}c@{}}Regular\\ Overlay\end{tabular}}} & \multicolumn{1}{c||}{\textbf{\begin{tabular}[c]{@{}c@{}}AS\\ Overlay\end{tabular}}} & \multicolumn{1}{c|}{\textbf{\begin{tabular}[c]{@{}c@{}}Image\\ Size\end{tabular}}} & \multicolumn{1}{c|}{\textbf{\begin{tabular}[c]{@{}c@{}}Bare \\ Metal\end{tabular}}} & \multicolumn{1}{c|}{\textbf{\begin{tabular}[c]{@{}c@{}}Regular \\ Overlay\end{tabular}}} & \multicolumn{1}{c||}{\textbf{\begin{tabular}[c]{@{}c@{}}AS \\ Overlay\end{tabular}}} & \multicolumn{1}{c|}{\textbf{\begin{tabular}[c]{@{}c@{}}Image\\ Size\end{tabular}}} & \multicolumn{1}{c|}{\textbf{\begin{tabular}[c]{@{}c@{}}Bare \\ Metal\end{tabular}}} & \multicolumn{1}{c|}{\textbf{\begin{tabular}[c]{@{}c@{}}Regular \\ Overlay\end{tabular}}} & \textbf{\begin{tabular}[c]{@{}c@{}}AS \\ Overlay\end{tabular}} \\ \hline 
8$\times$8 &0.39 &1.71 &  \multicolumn{1}{c||}{0.58}& 8$\times$8 & 0.043& 0.046 &  \multicolumn{1}{c||}{0.043}& 16$\times$16 &0.19 &0.58  & \multicolumn{1}{c||}{0.19} & 16$\times$16 &3.48 &4.35  &4.15  \\ \hline

16$\times$16 &3.05 & 13.65 & \multicolumn{1}{c||}{4.57} & 16$\times$16 & 0.113&  0.116&  \multicolumn{1}{c||}{0.113}& 32$\times$32 &0.756 &2.28 &  \multicolumn{1}{c||}{0.756}& 32$\times$32 & 13.68&17.1  &16.3  \\ \hline

32$\times$32 &24.29 &109.32  &  \multicolumn{1}{c||}{36.42}& 32$\times$32 &0.396 &0.400  &  \multicolumn{1}{c||}{0.396}& 64$\times$64 &3.03 &9.12  &  \multicolumn{1}{c||}{3.03}& 64$\times$64 & 54.72&68.40  &65.36  \\ \hline

64$\times$64 &194.20 &873.84  &  \multicolumn{1}{c||}{291.29}& 64$\times$64 &1.53 &1.54  &  \multicolumn{1}{c||}{1.53}& 128$\times$128 &12.13  &36.42 &  \multicolumn{1}{c||}{12.13}& 128$\times$128 & 218.52 & 273.15  &261.01  \\ \hline \hline
\end{tabular}
}
\end{table*}

\section{Experimental Observations}
\label{lab:experimentalObservation}
\subsection{Evaluation Platform and Setup}
\label{subsec:eval_platform}
For evaluation purposes, designs are implemented on a Xilinx Kintex UltraScale+ FPGA (xcku5p-ffvd900-2-i). Hardware generation is conducted with Vivado HLx Editions v2018.2, and RapidWright v2018.2.5-beta allows assembling hardware kernels. We ran Vivado, RapidWright, and the LLVM kernel mining on a computer equipped with an Intel Corei3-8130U CPU@2.20GHz$\times$4 and 8Gb of RAM.
We study image processing and matrix-based applications. Though kernels from applications can altogether be deployed on the AS-Overlay, we run applications individually with the purpose of assessing achievable performances, in particular: (1) global latency, (2) Fmax and productivity, (3) resource utilization, and (4) power consumption, when comparing AS-overlays to regular overlays and bare metal implementations. For each application, we design a $3\times3$ PEs overlay with the three flavors: \textit{\textbf{(i) Bare Metal}}: functions are implemented in HDL, embedded in PEs for multitasking, and compiled with Vivado.  \textit{\textbf{(ii) Regular Overlay}}: Each PE implements an ALU offering a dozen arithmetic and logic operations. The regular overlay is implemented in HDL and also compiled with Vivado. \textit{\textbf{(iii) AS-Overlay}}: Kernels are implemented into PEs using the design flow described in Figure \ref{fig:architecture}.

\subsection{Evaluation Results}
 Data are sent to the FPGA through a set of custom C functions as mentioned in the utilization flow of Figure \ref{fig:architecture}. Execution times recorded in Table \ref{tab:table2} come from placing the bare metal, regular overlay, and AS-Overlay implementations of each application alongside a MicroBlaze CPU with a $300$MHz global clock.
 \begin{wrapfigure}[12]{r}{0.26\textwidth}
\begin{center}
 \vspace{-10pt}
\hspace{-13pt}
\includegraphics[width=0.28\textwidth]{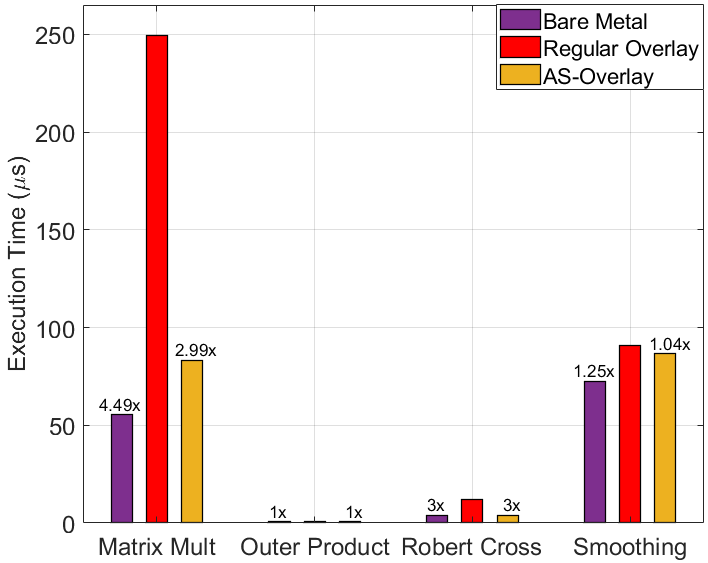}
\end{center}
\vspace{-10pt}
\caption{\label{fig:avg_execution_time} Execution Improvement}
\end{wrapfigure}
It shows that AS-Overlays can effectively compete with bare metal implementations in several test cases. The bare metal nevertheless outperforms AS-Overlays on the image smoothing and matrix multiplication because of the additional clock cycles introduced by the RapidWright application. In fact, to ensure timing closure when integrating kernels within the AS-Overlay fixed sections (PE architecture + interconnect), FFs are injected on the datapath after each operation. Table \ref{tab:table2} also shows that AS-Overlays compute faster that regular overlays when clocked with identical frequency. Figure \ref{fig:avg_execution_time} actually presents about $3$$\times$ improved execution time when averaging all the execution times of the tested applications. This performance gain is amplified by the Fmax study as higher clocked circuits can significantly reduce execution times. To carry out Fmax and productivity studies summarized in Table \ref{tab:table5}, for each application, we introduced a Phase-Locked Loop generating a $300$MHz (requesting higher frequencies like $400$MHz or $500$MHz returned negative slacks too high in some of the bare metal implementations) clock on each flavor of overlay. 
 The idea was to observe the maximum frequency and how long would the compilation take. As first observation, AS-Overlays can achieve up to $1.47$$\times$ improved Fmax compared to regular overlays on tested applications (the AS-Overlay tops at $447$MHz while the regular overlay caps at $304$MHz). This is caused by general-purpose ALUs of regular overlays that contain several muxes introducing substantial delays on datapaths. On the other hand, bare metal implementations achieved higher Fmax compared to AS-Overlays on outer product and Robert cross filter. It comes down to an observation made in \cite{lavin2018rapidwright}: vendor tools such as Vivado often produce high performance results for small modules of a design. In this case of figure, outer product and Robert cross are respectively a set of independent multiplications, and subtractions followed by comparisons, which gives to bare metal a $1.09$$\times$ Fmax advantage over AS-Overlays. That advantage is nevertheless lost on more complex functions such as image smoothing (the AS-Overlay achieved a $1.33$$\times$ higher Fmax), which computes the average of adjacent pixels, highlighting the benefits of using the RapidWright pre-implemented flow as smaller modules can be pre-implemented to achieve maximum frequency, and later be assembled with minimal QoR loss. Reported compilation times show that Kernel netlist generation and loading within PE black-boxes with the RapidWright application, outperforms up to $7$$\times$ Vivado synthesis both in Regular overlays and bare metal implementations. Table \ref{tab:table5} finally demonstrates that the proposed AS-overlay generation flow can provide up to $20$$\times$ productivity improvement over regular overlays on tested benchmarks.
\begin{figure}[]
	\centering
    \subfloat[ Number of Look-Up Tables]{
		\label{fig:luts}
		\includegraphics[width=0.2276\textwidth]{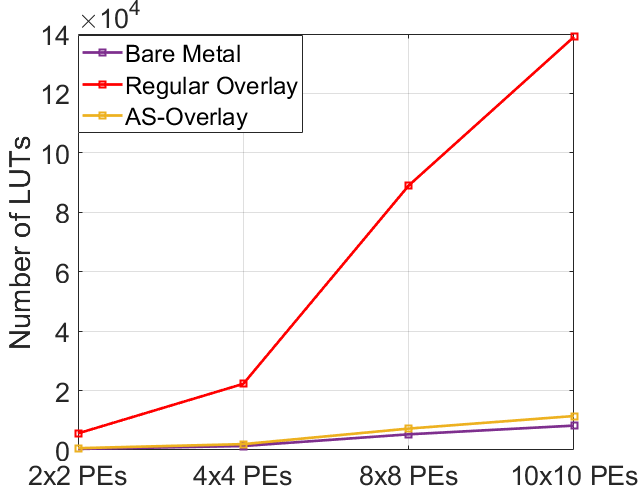}
	}		
	\hspace{0.0cm}
	\subfloat[Number of Flip-Flops]{
		\label{fig:ffs}
		\includegraphics[width=0.2276\textwidth]{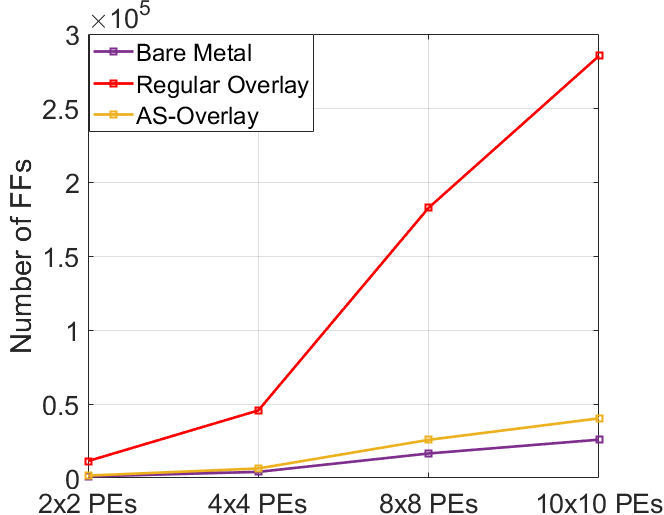}	}
  
    \vspace{-0.2cm}	
		
    \subfloat[ Number of DSP Blocks]{
		\label{fig:dsps}
		\includegraphics[width=0.2276\textwidth]{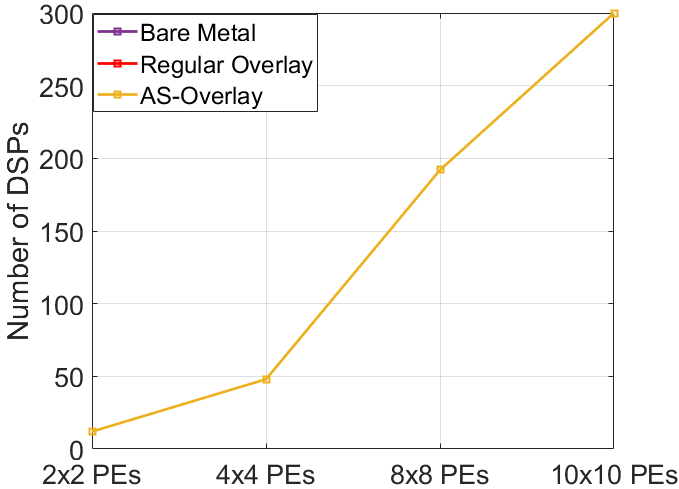}
	}		
	\hspace{0.0cm}
	\subfloat[Number of Block RAMs]{
		\label{fig:brams}
		\includegraphics[width=0.2276\textwidth]{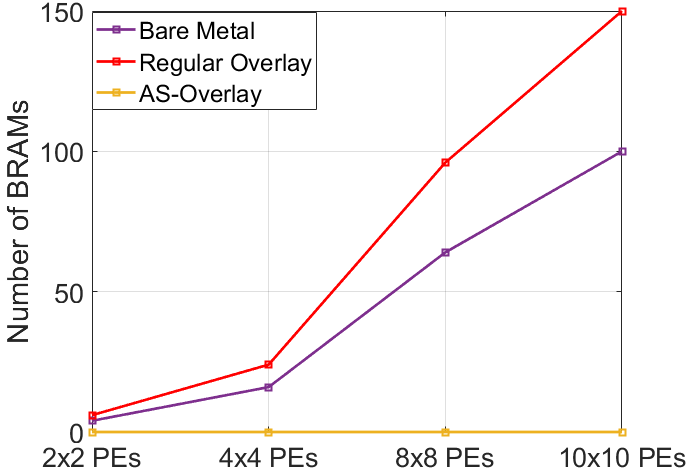}	}
 	\caption{FPGA Resource Utilization}
 	\label{fig:resourceUtilization} 
 	\vspace{-5pt}
\end{figure}

One way to load hardware kernels into PE black-boxes could have been to use the Vivado \textit{read\_checkpoint} TCL command in place of the RapidWright API like we did in the proposed approach. Vivado loading nevertheless outcomes in higher time and memory overhead as shown in Table \ref{tab:tab6}: Vivado in TCL mode or with the graphical user interface (GUI) incurs higher time penalty and RAM utilization than doing the same operation from RapidWright. While the RapidWright application uses few hundreds of Megabytes of the RAM on the testing computer, and loads kernels in about $2$ seconds, Vivado launched both with the GUI and the command line interface (CLI), uses about a Gigabyte of RAM and requires up to $6.19$ seconds to complete loading hardware kernels. This observation justifies why we only use Vivado for the placement and routing.
\vspace{-2pt}
\begin{center}
\begin{table}[]
\scriptsize
\center
\caption{Kernel Loading Time \& Memory Usage}
\label{tab:tab6}
\begin{tabular}{c|c||c|c|c|c}
\hline \hline
\multicolumn{2}{c||}{} & \textbf{\begin{tabular}[c]{@{}c@{}}Matrix \\ Mult\end{tabular}} & \textbf{\begin{tabular}[c]{@{}c@{}}Outer \\ Product\end{tabular}} & \textbf{\begin{tabular}[c]{@{}c@{}}Robert \\ Cross\end{tabular}} & \textbf{Smoothing} \\ \hline \hline
\multicolumn{2}{c||}{\textbf{\begin{tabular}[c]{@{}c@{}}RapidWright \\ Loading\end{tabular}}} & \begin{tabular}[c]{@{}c@{}} 2.16 s \\ \textbf{215.2 MB}\end{tabular} & \begin{tabular}[c]{@{}c@{}} 2.07 s \\ \textbf{129.6 MB}\end{tabular} & \begin{tabular}[c]{@{}c@{}} 2.13 s \\ \textbf{141.2 MB}\end{tabular} & \begin{tabular}[c]{@{}c@{}} 2.05 s \\ \textbf{142.2 MB}\end{tabular} \\ \hline
\multirow{2}{*}{\textbf{\begin{tabular}[c]{@{}c@{}}Vivado \\ Loading\end{tabular}}} & \rotatebox[origin=c]{90}{\textbf{ GUI }} & \begin{tabular}[c]{@{}c@{}} 5.62 s \\ \textbf{1.4 GB}\end{tabular} & \begin{tabular}[c]{@{}c@{}} 5.10 s \\ \textbf{1.4 GB}\end{tabular} & \begin{tabular}[c]{@{}c@{}} 5.32 s \\ \textbf{1.4 GB}\end{tabular} & \begin{tabular}[c]{@{}c@{}} 6.19 s \\ \textbf{1.5 GB}\end{tabular} \\ \cline{2-6} 
 & \rotatebox[origin=c]{90}{\textbf{ CLI }} & \begin{tabular}[c]{@{}c@{}} 2.19 s \\ \textbf{925.3 MB}\end{tabular} & \begin{tabular}[c]{@{}c@{}} 2.12 s \\ \textbf{924.2 MB}\end{tabular} & \begin{tabular}[c]{@{}c@{}} 2.18 s \\ \textbf{923.1 MB}\end{tabular} & \begin{tabular}[c]{@{}c@{}} 2.94 s \\ \textbf{936.1 MB}\end{tabular} \\ \hline \hline
\end{tabular}
\end{table}
\end{center}
\vspace{-10pt}

Figure \ref{fig:resourceUtilization} summarizes the utilization of FPGA resources only for the matrix multiplication (because of page limitation, it was not possible to present the same study for each of the tested applications) as an illustration of how the fabric is progressively occupied as the number of PEs is scaled up. In general, the total amount of resources used by AS-Overlays is close to that of the bare metal, and both far below regular overlays. Figure \ref{fig:dsps} nevertheless displays the same number of DSPs (the purple, red, and yellow lines are superimposed, so only the yellow line is visible) simply because PEs on the three platforms implement only one multiplier using 4 DSP48E2s.

\begin{figure}[h]
\centering
\vspace{-10pt}
\begin{minipage}[t]{4cm}
  \centering
  \includegraphics[scale=0.23]{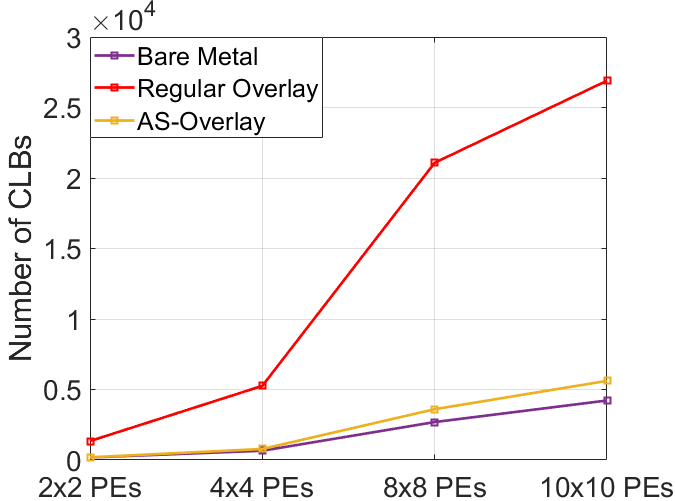}
  \caption{CLB Spreading}
  \label{fig:clb_spreading}
\end{minipage}
\hspace{0.2cm}
\begin{minipage}[t]{4cm}
  \centering
  \includegraphics[scale=0.21]{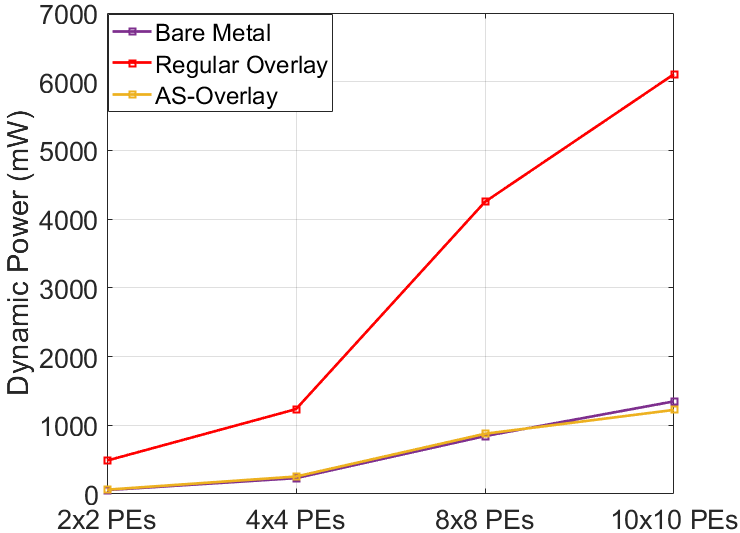}
  \caption{Power Consumption}
  \label{fig:powerConsumption}
\end{minipage}
\end{figure}
Pre-implementing basic functions from LLVM IR also have the potentiality of reducing resource utilization as illustrated in Figure  \ref{fig:brams}: Vivado optimizes individual hardware implementation from LLVM IR without BRAM insertion while adding such resources when compiling bare metal and regular overlays, which translates into a higher power consumption after $8\times8$ PEs (see Figure \ref{fig:powerConsumption}): with $10\time10$ PEs, the bare metal uses $1348$mW while the AS-Overlay consumes $1224$mW, about $1.10\times$ less power. Figure \ref{fig:ffs} reports a higher utilization of FFs in AS-Overlays as opposed to bare metal. This is due to FFs insertion during datapath regularization in addition to input, config and output registers from the PE architecture (check Figure \ref{fig:pe}). While injecting FFs on the datapath as explained in section \ref{sub:datapath_regularization} do not incur delays in kernel execution, it obviously increases the number of FFs used depending of the structure of kernel DFGs. This is nevertheless a price to pay to ensure the correctness of results produced by hardware kernels.
\vspace{-20pt}
\begin{center}
\begin{table}[h]
\scriptsize
\caption{Productivity Analysis \& Maximum Frequency}
\label{tab:table5}
\begin{tabular}{c|c|c|c|c|c}
\hline \hline
\multicolumn{2}{c||}{\multirow{2}{*}{\textbf{\begin{tabular}[c]{@{}c@{}} 3$\times$3 Overlay\\ Generation Flow \end{tabular}}}} & \multicolumn{4}{c}{\textbf{Applications}} \\ \cline{3-6} 
\multicolumn{2}{c||}{} & \multicolumn{1}{c|}{\textbf{\begin{tabular}[c]{@{}c@{}}Matrix \\ Mult\end{tabular}}} & \multicolumn{1}{c|}{\textbf{\begin{tabular}[c]{@{}c@{}}Outer \\ Product\end{tabular}}} & \multicolumn{1}{c|}{\textbf{\begin{tabular}[c]{@{}c@{}}Robert \\ Cross\end{tabular}}} & \multicolumn{1}{c}{\textbf{\begin{tabular}[c]{@{}c@{}}Smoothing\end{tabular}}} \\ \hline \hline
\multirow{6}{*}{\rotatebox[origin=c]{90}{\textbf{\begin{tabular}[c]{@{}c@{}}Bare Metal\end{tabular}}}} & \multicolumn{1}{l||}{\textit{Synthesis}} & 26 &17  &35  &35  \\ \cline{2-6} 
 & \multicolumn{1}{l||}{\textit{Optimization}} &  5 & 2 &6  &34  \\ \cline{2-6} 
 & \multicolumn{1}{l||}{\textit{Placement}} &  28& 23 & 44 &85  \\ \cline{2-6} 
 & \multicolumn{1}{l||}{\textit{Routing}} &  68& 59 & 55 & 1064 \\ \cline{2-6} 
 & \multicolumn{1}{l||}{\textbf{Total (Seconds)}} &\textbf{127}  & \textbf{101} & \textbf{140} &\textbf{1218}  \\ 
 & \multicolumn{1}{l||}{\textbf{Fmax (MHz)}} & \textbf{365} & \textbf{488} & \textbf{348} & \textbf{231}  \\ \hline \hline

 \multirow{3}{*}{\rotatebox[origin=c]{90}{\textbf{\begin{tabular}[c]{@{}c@{}}Regular \end{tabular}}}} 
 & \multicolumn{1}{l||}{\textbf{Vivado Flow $\rightarrow$}} &  \begin{tabular}[c]{@{}c@{}}\textit{\textbf{Synth.}} \\ 40 \end{tabular}& \begin{tabular}[c]{@{}c@{}}\textit{\textbf{Opt.}} \\ 27 \end{tabular}  & \begin{tabular}[c]{@{}c@{}}\textit{\textbf{Place.}} \\ 111 \end{tabular} &\begin{tabular}[c]{@{}c@{}}\textit{\textbf{Routing}} \\ 1457 \end{tabular}  \\ \cline{2-6}
 & \multicolumn{1}{l||}{\textbf{Total (Seconds)}}&\multicolumn{4}{c}{\textbf{1635 (27 minutes 15 seconds)}}     \\ 
 & \multicolumn{1}{l||}{\textbf{Fmax (MHz)}} &\multicolumn{4}{c}{\textbf{304}}  \\ \hline  \hline
 
\multirow{7}{*}{\rotatebox[origin=c]{90}{\textbf{\begin{tabular}[c]{@{}c@{}} AS-Overlay \end{tabular}}}} & \multicolumn{1}{l||}{\textit{\begin{tabular}[c]{@{}l@{}}Kernel Gen.\end{tabular}}} & 3.89 & 3.48  &3.55  &4.34  \\ \cline{2-6} 
 & \multicolumn{1}{l||}{\textit{\begin{tabular}[c]{@{}l@{}}Kernel Load.\end{tabular}}} & 2.16  & 2.07 &2.13  &2.05  \\ \cline{2-6} 
 & \multicolumn{1}{l||}{\textit{Optimization}} & 5 &  4&3  &33  \\ \cline{2-6} 
 & \multicolumn{1}{l||}{\textit{Placement}} & 46 &  19&24  & 83 \\ \cline{2-6} 
 & \multicolumn{1}{l||}{\textit{Routing}} &  65& 54  &117  &646  \\ \cline{2-6} 
 & \multicolumn{1}{l||}{\textbf{Total (Seconds)}} & \textbf{122.05} &  \textbf{82.55}& \textbf{149.68} &\textbf{768.39}  \\ 
 & \multicolumn{1}{l||}{\textbf{Fmax (MHz)}} & \textbf{435} & \textbf{447} &  \textbf{318}&\textbf{308}  \\ \hline \hline

\end{tabular}
\end{table}
\end{center}
\vspace{-25pt}

We also assess the overall use of the FPGA layout. Without defining pblock constraints on the designs, we study how Vivado spreads circuits across Configurable Logic Blocks (CLBs) on the FPGA. This provides a good measurement of how much space remains available on the fabric (see Figure \ref{fig:clb_spreading}). For $10\times10$ PEs, the regular overlay is spread over $26901$ CLBs, about $99.19$\% of available resources on the Kintex UltraScale+ \cite{ultraScale}, making it impossible to load any other design on the chip. On the other hand, the bare metal and AS-Overlay respectively use $15.56$\% and $20.70$\% of available CLBs, leaving enough room to fit the domain specific implementation alongside other design modules on a single chip. Table \ref{tab:table1} quantifies the maximum number of inputs and outputs of kernel identified on used benchmarks, and the overhead associated with the kernel mining. Adding the pass for kernel detection within LLVM only incurs additional compilation time in the magnitude of milliseconds. The smoothing recorded the highest number of inputs. With each datapath over $32$-bits, handling $10\times32$-bits is not an issue on modern FPGAs.
\begin{center}
\begin{table}[]
\scriptsize
\center
\caption {Kernel I/O \& Compilation Time Comparison} \label{tab:table1}
\begin{tabular}{c|c|c|c|c}
\hline \hline
\textbf{Applications} & \multicolumn{1}{c|}{\textbf{\begin{tabular}[c]{@{}c@{}}Kernel\\ Inputs\end{tabular}}} & \multicolumn{1}{c||}{\textbf{\begin{tabular}[c]{@{}c@{}}Kernel\\ Outputs\end{tabular}}}  & \multicolumn{1}{c|}{\textbf{\begin{tabular}[c]{@{}c@{}}LLVM \\ Compilation \end{tabular}}}  & \multicolumn{1}{c}{\textbf{\begin{tabular}[c]{@{}c@{}}LLVM + \\ Kernel Mining \end{tabular}}}    \\ \hline \hline
\begin{tabular}[c]{@{}c@{}}Matrix Mult\end{tabular} & 3 & \multicolumn{1}{c||}{1} & 4.12s & 4.59s (1.1$\times\downarrow$)\\ \hline
\begin{tabular}[c]{@{}c@{}}Outer Product\end{tabular} & 2 & \multicolumn{1}{c||}{1} & 0.048s & 0.12s (2.5$\times\downarrow$)\\ \hline
\begin{tabular}[c]{@{}c@{}}Robert Cross \\\end{tabular} &4  &  \multicolumn{1}{c||}{1}& 0.15s & 0.30s (2$\times\downarrow$)\\ \hline
\begin{tabular}[c]{@{}c@{}}Smoothing\end{tabular} &10  & \multicolumn{1}{c||}{1}& 0.16s & 0.26s (1.6$\times\downarrow$)\\ \hline \hline
\end{tabular}
\vspace{-5pt}
\end{table}
\end{center}

Ma et al. \cite{ma2015run} reported a productivity improvement between $170\times$ and $214\times$, which unfortunately only accounts synthesis (no details are provided on placement and routing time). Further, they did not discuss data size. Finally, they used benchmarks, a Vivado version, and an FPGA different from ours, with no information on the characteristics of the machine used for compilation. Similar observations can be made on other works from section \ref{lab:relatedWork}. Overall, establishing a fair comparison of results with previous work is particularly challenging because of the impossibility of reproducing identical experimental environments.
\vspace{-5pt}
\section{Conclusion}
\label{lab:conclusion}
\vspace{-2pt}
In this paper, we presented an approach aiming the automatic generation of Application-Specific FPGA Overlays for data flow applications capable of providing bare metal performances. The approach extracts kernels from applications at compile time, and automatically builds accelerators tailored for the application needs. Experimental evaluations demonstrated the viability of our approach with significant productivity improvement, power consumption reduction, and lower execution time over regular FPGA overlays. Future work will investigate the replicability feature of RapidWright coupled with LLVM code instrumentation to build more efficient FPGA accelerators.

\bibliographystyle{./IEEEtran}
\bibliography{fpl-2019-ds-overlay}

\end{document}